# MHD flow and heat transfer due to a linearly stretching sheet with induced magnetic field: Exact solution

Tarek M. A. El-Mistikawy

Dept. Eng. Math. & Phys., Faculty of Engineering, Cairo University, Giza 12211, Egypt

E-mail: elmistikawy@eng.cu.edu.eg

**Abstract**

The solution for the magnetohydrodynamic flow, due to a linearly stretching sheet, has a simple form for the velocity field, with a companion simple form for the induced magnetic field. The associated thermal problem, including viscous dissipation and Joule heating, involves three temperature constituents, the solutions for two of which are obtained in terms of Kummer's function. The solution for the third temperature constituent is obtained in a convergent series form.

1. **Introduction**

Navier-Stokes equations are inherently nonlinear due to the presence of convection terms. When convection is absent, the equations become linear and exact (closed form ) solutions may be possible; e.g. Couette and Poiseuille flows [1].

With convection present, exact solutions to Navier-Stokes equations are rare. This is to the extent that reduction of order through self-similarity is regarded as exact solution [2,3].

Crane's simple exact solution [4] to the problem of the flow due to a linearly stretching sheet is one such rare solution. This invited researchers to apply the same form of solution to other related situations, which proved successful in cases involving surface feed [5], velocity slip [6], magnetohydrodynamic (MHD) flow (under the assumption of negligible induced magnetic field) [7], or combinations thereof [8].

The thermal problem associated with Crane's flow is a linear one. It was solved by Grupka and Bobba [9] neglecting viscous dissipation and streamwise heat diffusion. Neglecting Joule heating as well, Liu [10] solved the thermal problem associated with Andersson's MHD flow [7], in cases of linear surface temperature and linear heat flux.

It is shown here that the solution for the MHD flow with induced magnetic field has Crane's simple form for the velocity field, with a companion simple form for the induced magnetic field. Both solutions tend to the leading order solutions obtained in [11], as the magnetic Prandtl number diminishes. The associated thermal problem, including viscous dissipation and Joule heating, involves quadratic streamwise variation in temperature with three temperature constituents, the solutions for two of which are obtained in terms of Kummer's functions [12]. The solution for the third temperature constituent is obtained in convergent series form.

## 2. Mathematical model

An electrically conducting, incompressible, Newtonian fluid is driven by a non-conducting non-porous sheet, which is stretching linearly in the $x$-direction with rate $\omega$. At the surface, we consider cases of prescribed temperature or heat flux. In the farfield, the fluid is essentially quiescent under pressure $p = p_\infty$ and temperature $T = T_\infty$, and is permeated by a stationary magnetic field of uniform strength $B$ in the transverse $y$-direction. The velocity components in the $(x,y)$ directions are $(u,v)$ and the corresponding induced magnetic field components are $(q,s)$. Constants are the fluid density $\rho$, kinematic viscosity $\vartheta$, electric conductivity $\sigma$, magnetic permeability $\mu$, specific heat $c$, and thermal conductivity $k$.

The problem admits the similarity transformations [13]

$$x = \sqrt{\vartheta/\omega}\,\chi,\ y = \sqrt{\vartheta/\omega}\,\eta,\ v = -\sqrt{\vartheta\omega}\,f(\eta),\ u = \sqrt{\vartheta\omega}\,\chi f' \qquad (1\text{a,b,c,d})$$

$$s = B\sigma\mu\vartheta g(\eta),\ q = -B\sigma\mu\vartheta\chi g' \qquad (1\text{e,f})$$

$$p = p_\infty - \rho\omega\vartheta[f' + \tfrac{1}{2}f^2(\eta) - \tfrac{1}{2}f^2(\infty)] - \tfrac{1}{2}B^2\sigma^2\mu\vartheta^2\chi^2 g'^2 \qquad (1\text{g})$$

$$T = T_\infty + \frac{\vartheta\omega}{c}[\theta_0(\eta) + \chi\theta_1(\eta) + \chi^2\theta_2(\eta)] \qquad (1\text{h})$$

where primes denote differentiation with respect to the similarity coordinate $\eta$.

The governing fluid flow continuity and Navier-Stokes equations, electromagnetic Maxwell's and Ohm's equations, and energy equation including heat dissipation and Joule heating take the similarity form [13]

$$f''' + ff'' - f'^2 = \beta[\text{P}_\text{m}(gg'' - g'^2) + g''] \qquad (2)$$

$$g'' = f' + \text{P}_\text{m}(gf' - fg') \qquad (3)$$

$$\theta_2'' + \text{Pr}(f\theta_2' - 2f'\theta_2) = -\text{Pr}\,(\beta g''^2 + f''^2) \qquad (4)$$

$$\theta_1'' + \text{Pr}(f\theta_1' - f'\theta_1) = 0 \qquad (5)$$

$$\theta_0'' + \text{Pr}f\theta_0' = -2\theta_2 - 4\text{Pr}\,f'^2 \qquad (6)$$

where $\text{P}_\text{m} = \sigma\mu\vartheta$ is the magnetic Prandtl number, $\beta = \sigma B^2/\rho\omega$ is the magnetic interaction number, $\text{Pr} = \rho\vartheta c/k$ is the Prandtl number.

In the right-hand-sides of Eqs. (4) and (6), the terms involving $f''^2$ and $f'^2$ are due to viscous dissipation, the term involving $g''^2$ is due to Joule heating, and the term involving $\theta_2$ is due to the streamwise heat diffusion.

The boundary conditions, in the absence of velocity and thermal slip, are [13]

$$f(0) = 0, f'(0) = 1, f'(\infty) = 0 \tag{7}$$

$$g(\infty) = 0, g'(\infty) = 0 \tag{8}$$

$$\theta_2(0) = \Theta_2 \text{ or } \theta_2'(0) = -Q_2, \theta_2(\infty) = 0 \tag{9}$$

$$\theta_1(0) = \Theta_1 \text{ or } \theta_1'(0) = -Q_1, \theta_1(\infty) = 0 \tag{10}$$

$$\theta_0(0) = \Theta_0 \text{ or } \theta_0'(0) = -Q_0, \theta_0(\infty) = 0 \tag{11}$$

where $\Theta_0$, $\Theta_1$ and $\Theta_2$ are prescribed constituents of the surface temperature, while $Q_0$, $Q_1$ and $Q_2$ are prescribed constituents of the heat flux from the surface.

Conditions (8) on $g$ translate the physical requirement of the absence of the current density in the farfield, and indicate that $B$ stands for the farfield total magnetic field imposed and induced [11].

3. **Exact solution for the velocity and magnetic fields**

We attempt solutions of the form

$$f = (1 - e^{-\gamma\eta})/\gamma \tag{12}$$

$$g = \lambda e^{-\gamma\eta} \tag{13}$$

which satisfy the boundary conditions, provided $\gamma > 0$.

Substitution in Eqs. (2) and (3) gives the following relations between $\gamma$ and $\lambda$.

$$\gamma^2 - 1 = \beta\gamma^2\lambda \tag{14}$$

$$\gamma^2\lambda = 1 + P_m\lambda \tag{15}$$

Elimination of $\lambda$ results in

$$\gamma^4 - (\beta + 1 + P_m)\gamma^2 + P_m = 0 \tag{16}$$

with the solutions

$$\gamma^2 = \tfrac{1}{2}[(\beta + 1 + P_m) \pm \sqrt{(\beta + 1 + P_m)^2 - 4P_m}] \tag{17}$$

The plus sign is chosen, so that $\gamma^2 \neq 0$ when $P_m = 0$, leading to

$$\gamma = \sqrt{\tfrac{1}{2}[(\beta + 1 + P_m) + \sqrt{(\beta + 1 + P_m)^2 - 4P_m}]} \tag{18}$$

and, consequently,

$$\lambda = 2[(\beta + 1 - P_m) + \sqrt{(\beta + 1 + P_m)^2 - 4P_m}]^{-1} \tag{19}$$

Note that both $\gamma$ and $\lambda$ are real, since the discriminant $(\beta + 1 + P_m)^2 - 4P_m$ manipulates to $(\beta + 1 - P_m)^2 + 4\beta P_m > 0$. Moreover, $\gamma > 1$ and $\lambda > 0$.

The solution for $f$ reduces to Andersson's solution [7] in the case of negligible induced magnetic field ($P_m = 0$), and to Crane's solution [4] in the hydrodynamic case ($\beta = P_m = 0$). In either case, Eq. (3) merely defines $g'' = f'$.

4. **Exact solutions for the temperature constituents**

Upon substitution for $f$ and $g$, Eqs. (4)-(6) combine as

$$\phi_n'' + r(1 - e^{-\zeta})\phi_n' - nre^{-\zeta}\phi_n = \Lambda_n \tag{20}$$

where $r = \Pr/\gamma^2$, $\zeta = \gamma\eta$, $\phi_n(\zeta) = \theta_n(\eta)/\gamma^n$, and $n = 2, 1$ or $0$, with

$$\Lambda_2 = -r(1 + \beta\gamma^2\lambda^2)e^{-2\zeta} \tag{21}$$

$$\Lambda_1 = 0 \tag{22}$$

$$\Lambda_0 = -2\phi_2 - 4re^{-2\zeta} \tag{23}$$

and a prime denoting differentiation with respect to $\zeta$. Henceforth, it is assumed that $r > 0$ is not an integer.

It is realized that Eq. (20) can be transformed into Kummer's equation [12] with a source term. The solution can, thus, be expressed as the sum of a particular part due to the source term and a homogeneous part in terms of Kummer's function

$$M(a,b,z) = \sum_{m=0}^{\infty} \frac{a^{(m)}}{m!\, b^{(m)}} z^m$$

where $a^{(m)} = a(a+1)\ldots(a+m-1)$ with $a^{(0)} = 1$.

In particular, the solution for $\phi_2$ takes the form

$$\phi_2(\zeta) = c_0 + c_1 e^{-\zeta} + A_2 M(-2, 1-r, -re^{-\zeta}) \\ + B_2 e^{-r\zeta} M(-2+r, 1+r, -re^{-\zeta}) \tag{24}$$

where $A_2$ and $B_2$ are arbitrary constants of integration, while $c_0$ and $c_1$ are undetermined coefficients of the particular solution.

For the satisfaction of the farfield condition (9c), $A_2 = -c_0$. On the other hand, the surface condition (9a) gives

$$\Theta_2/\gamma^2 = c_0[1 - M(-2, 1-r, -r)] + c_1 + B_2 M(-2+r, 1+r, -r) \tag{25a}$$

while condition (9b) gives

$$Q_2/\gamma^2 = \gamma c_1 - \gamma c_0 \frac{2r}{1-r} M(-1, 2-r, -r) \\ + B_2 \gamma r [M(-2+r, 1+r, -r) + \frac{2-r}{1+r} M(-1+r, 2+r, -r)] \tag{25b}$$

Either Eq. (25a) or Eq. (25b) can be solved for $B_2$ in terms of $c_0$ and $c_1$. These two coefficients are determined by the satisfaction of Eq. (20) for $n=2$, which gives

$$c_1 = 1 + \beta\gamma^2 \lambda^2 \text{ and } c_0 = \frac{1-r}{2r} c_1 \tag{26}$$

The solution for $\phi_1$ is

$$\phi_1(\zeta) = B_1 e^{-r\zeta} M(-1+r, 1+r, -re^{-\zeta}) \tag{27}$$

where $B_1$ satisfies

$$\Theta_1/\gamma = B_1 M(-1+r, 1+r, -r) \tag{28a}$$

or

$$Q_1/\gamma = B_1 \gamma r [M(-1+r, 1+r, -r) + \frac{1-r}{1+r} M(r, 2+r, -r)] \tag{28b}$$

The solution for $\phi_0$ is

$$\phi_0(\zeta) = B_0 e^{-r\zeta} M(r, 1+r, -re^{-\zeta}) + \varphi_0(\zeta) \tag{29}$$

where the particular solution $\varphi_0$ is given by

$$\varphi_0 = -[4c_0 \frac{(-r)^2}{(-r)^{(2)}} + 2c_1] \sum_{k=1}^{\infty} \frac{(-r)^k}{k(-r)^{(k+1)}} e^{-k\zeta}$$

$$+ [2c_0 \frac{(-r)^3}{(-r)^{(3)}} - 4r] \sum_{k=2}^{\infty} \frac{(-r)^{k-2}}{k(2-r)^{(k-1)}} e^{-k\zeta}$$

$$+ 2B_2 e^{-r\zeta} \sum_{k=1}^{\infty} \frac{(-r)^k}{k!(r+k)} e^{-k\zeta} \sum_{m=1}^{k} \frac{r(3r+3m-7)+(m-2)^2}{(r+m)(r+m-1)(r+m-2)}$$

$$+ 2B_2 \zeta e^{-r\zeta} \sum_{k=0}^{\infty} \frac{(-r)^k}{k!(r+k)} e^{-k\zeta} \tag{30}$$

and $B_0$ satisfies

$$\Theta_0 = B_0 M(r, 1+r, -r) + \varphi_0(0) \tag{31a}$$

or

$$Q_0 = B_0 \gamma r \left[ M(r, 1+r, -r) - \frac{r}{1+r} M(1+r, 2+r, -r) \right] - \gamma \varphi_0'(0) \tag{31b}$$

The convergence of the summations in Eq. (30) is established in Appendix A.

It is noted that, when the $\Theta_n$'s are given, the $Q_n$'s in Eqs. (25b), (28b) and (31b) determine the constituents of the rate of heat transfer from the surface to the fluid. Conversely, when the $Q_n$'s are given, the $\Theta_n$'s in Eqs. (25a), (28a) and (31a) determine the constituents of the surface temperature. Sample results are given in Appendix B.

Note: In Appendix C, simplifications to some expressions are obtained, using properties of Kummer's function. In Appendix D, the exact solutions are extended to include cases of surface feed, velocity slip and thermal slip.

## 5. Conclusion

It is shown that Crane's simple form of exact solution, for the hydrodynamic problem of the flow due to a linearly stretching sheet, extends to the current MHD problem with a companion simple form for the induced magnetic field. Moreover, an exact solution for the energy equation including viscous dissipation, Joule heating, and streamwise heat diffusion is obtained. The obtained exact solutions tend regularly to previously published solutions of degenerate problems [4,7,9,10].

The inclusion of the traditionally ignored physical processes of magnetic induction, viscous dissipation and Joule heating causes streamwise variations in the pressure and temperature. As Eq. (1g) indicates, the induced magnetic field instigates a favorable pressure gradient proportional to the streamwise coordinate $\chi$. Even when the surface temperature or heat flux is streamwise uniform, viscous dissipation and (or) Joule heating bring(s) about temperature variations proportional to $\chi^2$.

## References

[1] H. Shlichting, K. Gersten, Boundary-Layer Theory, 8th edition, Springer, Berlin, 2003, pp. 101-103.

[2] C.Y. Wang, Exact solutions of the steady-state Navier–Stokes equations, Annual Review of Fluid Mechanics 23 (1991) 159–177.

[3] C. Y. Wang, Review of similarity stretching exact solutions of the Navier–Stokes equations, European Journal of Mechanics B/Fluids 30 (2011) 475–479.

[4] L. J. Crane, Flow past a stretching plate, Journal of Applied Mathematics and Physics ZAMP 21 (1970) 645– 647.

[5] P. S. Gupta, A. S. Gupta, Heat and mass transfer on a stretching sheet with suction and blowing, Canadian Journal of Chemical Engineering 55 (1977) 744–746.

[6] H. I. Andersson, Slip flow past a stretching surface, Acta Mechanica 158 (2002) 121–125.

[7] H. I. Andersson, An exact solution of the Navier-Stokes equations for magnetohydrodynamic flow, Acta Mechanica 113 (1995) 241–244.

[8] T. Fang, J. Zhang, S. Yao, Slip MHD viscous flow over a stretching sheet – An exact solution, Communications in Nonlinear Science and Numerical Simulation 14 (2009) 3731–3737.

[9] L. J. Grubka, K. M. Bobba, Heat Transfer Characteristics of a Continuous, Stretching Surface With Variable Temperature, ASME Journal of Heat Transfer 107 (1985) 248–250.

[10] I-C Liu, A note on heat and mass transfer for a hydromagnetic flow over a stretching sheet, International Communications in Heat and Mass Transfer 32 (2005) 1075–1084.

[11] T. M. A. El-Mistikawy, MHD flow due to a linearly stretching sheet with induced magnetic field, Acta Mechanica 227 (2016) 3049–3053.

[12] A. B. Olde Daalhuis, NIST Digital Library of Mathematical Functions, chapter 13, 2010. URL: dlmf.nist.gov/13.2.

[13] T. M. A. El-Mistikawy, Heat transfer in MHD flow due to a linearly stretching sheet with induced magnetic field, Advances in Mathematical Physics 2018 (2018), Article ID 5686089.

**Appendix A**

For the summation

$$S = \sum_{k=1}^{\infty} \frac{(-r)^k}{k!\,(r+k)} e^{-k\zeta} \sum_{m=1}^{k} \frac{r(3r+3m-7)+(m-2)^2}{(r+m)(r+m-1)(r+m-2)}$$

which appears in Eq. (30), we note that

$$b_m = \frac{r(3r+3m-7)+(m-2)^2}{(r+m)(r+m-1)(r+m-2)}$$

and

$$b_m - b_{m+1} = \frac{(m-2)(m-3)+6r^2+4(m-3)r}{(r+m+1)(r+m)(r+m-1)(r+m-2)}$$

indicate that, for integer $m \geq 3$

$$b_m > b_{m+1} > 0$$

regardless of the value of $r$. Moreover,

$$\lim_{m \to \infty} b_m = 0$$

With

$$C_k = \sum_{m=3}^{k} b_m > 0$$

we have

$$\lim_{k \to \infty} \frac{C_{k+1}}{C_k} = \lim_{k \to \infty} \left(1 + \frac{b_{k+1}}{C_k}\right) = 1$$

and we can write

$$S = -\frac{(3r-1)}{(r+1)^2}e^{-\zeta} + \frac{r(3r-1)}{(r+2)^2}e^{-2\zeta} + \frac{2(3r-1)}{(r+2)r}\sum_{k=3}^{\infty}\frac{(-r)^k}{k!(r+k)}e^{-k\zeta} + \sum_{k=3}^{\infty}\frac{(-r)^k}{k!(r+k)}C_k e^{-k\zeta}$$

Thus, the infinite summations included in Eq. (30) have the forms

$$\sum_k \frac{(-r)^k}{k(-r)^{(k+1)}}e^{-k\zeta}, \sum_k \frac{(-r)^{k-2}}{k(2-r)^{(k-1)}}e^{-k\zeta}, \sum_k \frac{(-r)^k}{k!(r+k)}e^{-k\zeta}, \sum_k \frac{(-r)^k}{k!(r+k)}C_k e^{-k\zeta}$$

whose convergence can be easily established through the ratio test.

## Appendix B

The exact solutions for the thermal problem are used to produce sample numerical values, in order to demonstrate their usefulness.

In Table 1, we compare the present results to previously published results. The problem under consideration is one with linearly varying surface temperature ($\Theta_0 = 0$, $\Theta_1 = 1$, $\Theta_2 = 0$). The physical processes of viscous dissipation and streamwise diffusion are ignored in [9] and in [10], which ignores magnetic induction and Joule heating, as well. Only one heat transfer constituent, $Q_1$, is involved. Retaining the ignored processes, revives the other two constituents $Q_0$ and $Q_2$. Further, comparison with the numerically obtained results in [13] indicates the high accuracy of the numerical method adopted in [13]. This illustrates one benefit of exact solutions.

Table 1 Comparison of present results to literature results.

| Case | Ref. | $Q_0$ | $Q_1$ | $Q_2$ | $Q_0$ | $Q_1$ | $Q_2$ |
|---|---|---|---|---|---|---|---|
| $\beta = 0$, $P_m = 0$, $Pr = 0.72$ | [9] | 0 | 0.8086 | 0 | -1.49116 | 0.80863 | -0.25635 |
| $\beta = 1$, $P_m = 0$, $Pr = 0.7$ | [10] | 0 | 0.689699 | 0 | -1.52369 | 0.689699 | -0.59691 |
| $\beta = 1$, $P_m = 0.1$, $Pr = 0.72$ | [13] | -1.56122 | 0.69997 | -0.63757 | -1.56119 | 0.69997 | -0.63758 |
| | | | | | | Present | |

The effect of neglecting one or all of the abovementioned thermal processes is presented in Tables 2a,b for Case 1 when the surface is maintained at the freestream temperature, and in Tables 3a,b for Case 2 when the surface is thermally insulated. In either case, $\theta_1$ is governed by a homogeneous problem, obviating its contribution. Moreover, profiles for the temperature gradients $\theta_0'$ and $\theta_2'$ in Case 1, and for the temperature constituents $\theta_0$ and $\theta_2$ in Case 2 are shown in Figs. 1 and 2, respectively.

The following is reported.
- The contribution of the induced magnetic field is minor, even at the high value of the magnetic Prandtl number $P_m = 0.1$, invoked throughout.
- Heat generated within the fluid, through viscous friction and Joule heating, results in heat flux to the surface in Case 1, and in rise in surface temperature in Case 2.
- The streamwise diffusion, which is ignored in boundary layer formulations, is of appreciable effect, increasing as the magnetic interaction number $\beta$ increases and as the Prandtl number Pr decreases.
- Contributions of the constituent $\theta_2$ magnify downstream, being multiplied by $\chi^2$.

Results for Case 1: Constant Surface Temperature

Table 2 Effect of magnetic induction, viscous dissipation, Joule heating and streamwise diffusion.

a. $\text{Pr} = 0.72$

| Case | Neglecting All Effects | Neglecting Magnetic Induction | Neglecting Viscous Dissipation | Neglecting Joule Heating | Neglecting Streamwise Diffusion | Retaining All Effects | $\beta$ |
|---|---|---|---|---|---|---|---|
| $Q_0$ | 0 | -1.48466 | -0.02879 | -1.45805 | -1.19353 | -1.48684 | 0.1 |
| $Q_2$ | 0 | -0.29860 | -0.02997 | -0.27540 |  | -0.30537 |  |
| $Q_0$ | 0 | -1.54337 | -0.22394 | -1.33725 | -0.92158 | -1.56119 | 1 |
| $Q_2$ | 0 | -0.61143 | -0.22323 | -0.41435 |  | -0.63758 |  |
| $Q_0$ | 0 | -3.09839 | -1.29233 | -1.83132 | -0.42373 | -3.12365 | 10 |
| $Q_2$ | 0 | -2.15077 | -1.03905 | -1.13172 |  | -2.17077 |  |

b. $\text{Pr} = 7.0$

| Case | Neglecting All Effects | Neglecting Magnetic Induction | Neglecting Viscous Dissipation | Neglecting Joule Heating | Neglecting Streamwise Diffusion | Retaining All Effects | $\beta$ |
|---|---|---|---|---|---|---|---|
| $Q_0$ | 0 | -8.71366 | -0.02238 | -8.67228 | -8.4666 | -8.69466 | 0.1 |
| $Q_2$ | 0 | -1.62865 | -0.16367 | -1.50397 |  | -1.66764 |  |
| $Q_0$ | 0 | -7.42052 | -0.15854 | -7.22010 | -6.92583 | -7.37864 | 1 |
| $Q_2$ | 0 | -3.62323 | -1.32751 | -2.46403 |  | -3.79155 |  |
| $Q_0$ | 0 | -5.28842 | -0.77206 | -4.52084 | -3.67993 | -5.29290 | 10 |
| $Q_2$ | 0 | -16.14890 | -7.81008 | -8.50665 |  | -16.31674 |  |

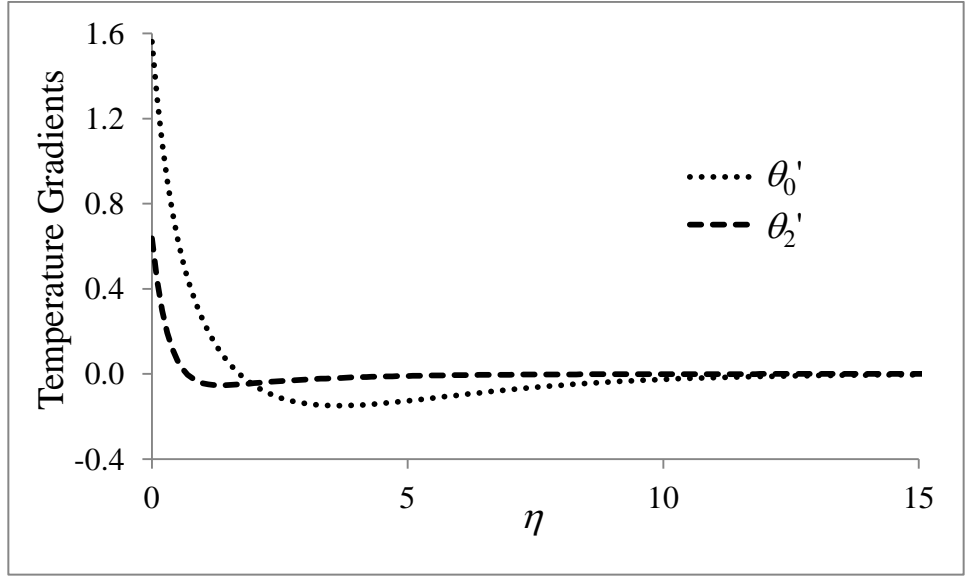

Fig. 1 Temperature Gradients. $\beta = 1$, $P_m = 0.1$, $Pr = 0.72$.

Results for Case 2: Thermally Insulated Surface

Table 3 Effect of magnetic induction, viscous dissipation, Joule heating and streamwise diffusion.

a. $Pr = 0.72$

| Case | Neglecting All Effects | Neglecting Magnetic Induction | Neglecting Viscous Dissipation | Neglecting Joule Heating | Neglecting Streamwise Diffusion | Retaining All Effects | $\beta$ |
|---|---|---|---|---|---|---|---|
| $\Theta_0$ | 0 | 4.38432 | 0.17596 | 4.25335 | 2.63645 | 4.42930 | 0.1 |
| $\Theta_2$ | 0 | 0.27786 | 0.0 2792 | 0.25659 | | 0.28452 | |
| $\Theta_0$ | 0 | 8.02607 | 2.10342 | 6.29289 | 2.38869 | 8.39631 | 1 |
| $\Theta_2$ | 0 | 0.62938 | 0.23097 | 0.42871 | | 0.65967 | |
| $\Theta_0$ | 0 | 157.40217 | 75.90184 | 84.75464 | 2.08323 | 160.65647 | 10 |
| $\Theta_2$ | 0 | 3.71306 | 1.79966 | 1.96016 | | 3.75982 | |

b. $Pr = 7.0$

| Case | Neglecting All Effects | Neglecting Magnetic Induction | Neglecting Viscous Dissipation | Neglecting Joule Heating | Neglecting Streamwise Diffusion | Retaining All Effects | $\beta$ |
|---|---|---|---|---|---|---|---|
| $\Theta_0$ | 0 | 4.69997 | 0.01960 | 4.67477 | 4.49464 | 4.69437 | 0.1 |
| $\Theta_2$ | 0 | 0.41139 | 0.04135 | 0.38001 | | 0.42136 | |
| $\Theta_0$ | 0 | 4.30634 | 0.15985 | 4.14265 | 3.84595 | 4.30251 | 1 |
| $\Theta_2$ | 0 | 0.93778 | 0.34401 | 0.63853 | | 0.98255 | |
| $\Theta_0$ | 0 | 5.79220 | 1.53860 | 4.29908 | 2.62325 | 5.83768 | 10 |
| $\Theta_2$ | 0 | 4.83854 | 2.34280 | 2.55175 | | 4.89454 | |

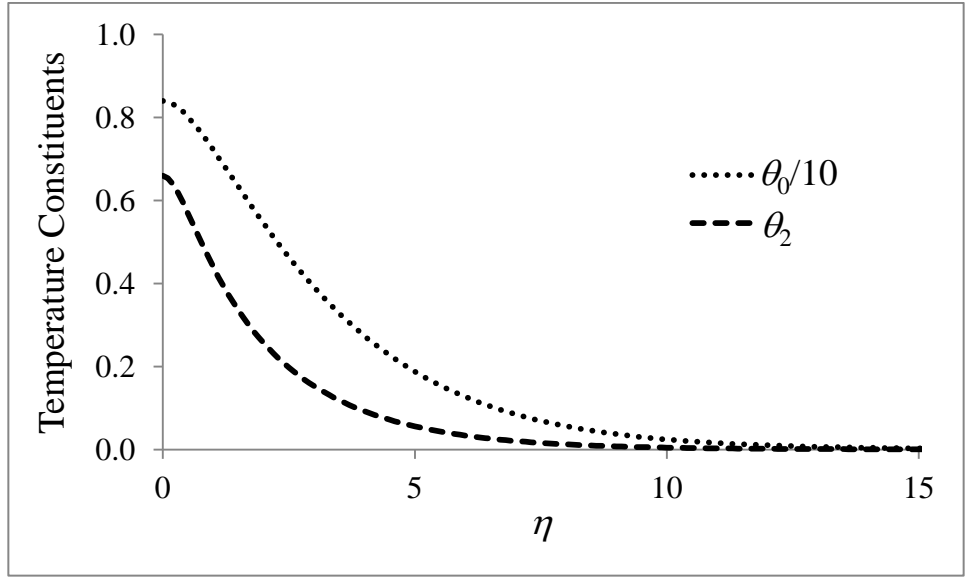

Fig. 2 Temperature Constituents. $\beta = 1$, $P_m = 0.1$, $Pr = 0.72$.

**Appendix C**

It is realized that Kummer's function $M[-2, 1-r, -re^{-\zeta}]$ expands as

$$M[-2, 1-r, -re^{-\zeta}] = 1 + \frac{2r}{1-r}e^{-\zeta} + \frac{r^2}{(1-r)(2-r)}e^{-2\zeta}$$

This simplifies the following formulas given the same equation number as the main text.

The solution for $\phi_2(\zeta)$ reduces to

$$\phi_2 = -(1+\beta\gamma^2\lambda^2)\frac{r}{2(2-r)}e^{-2\zeta} + B_2 e^{-r\zeta}M(-2+r, 1+r, -re^{-\zeta}) \quad (24a)$$

with the derivative

$$\phi_2' = (1+\beta\gamma^2\lambda^2)\frac{r}{2-r}e^{-2\zeta} - rB_2 e^{-r\zeta}[M(-2+r, 1+r, -re^{-\zeta}) \\ + \frac{2-r}{1+r}e^{-\zeta}M(-1+r, 2+r, -re^{-\zeta})] \quad (24b)$$

The surface condition (9a) gives

$$\Theta_2/\gamma^2 = -(1+\beta\gamma^2\lambda^2)\frac{r}{2(2-r)} + B_2 M(-2+r, 1+r, -r) \quad (25a)$$

while condition (9b) gives

$$Q_2/\gamma^2 = -(1+\beta\gamma^2\lambda^2)\frac{\gamma r}{2-r}$$
$$+B_2\gamma r[M(-2+r,1+r,-r)+\frac{2-r}{1+r}M(-1+r,2+r,-r)] \quad (25b)$$

The particular solution for $\phi_0$ reduces to

$$\varphi_0 = r[(1+\beta\gamma^2\lambda^2)\frac{1}{2-r} - 4]\sum_{k=2}^{\infty}\frac{(-r)^{k-2}}{k(2-r)^{(k-1)}}e^{-k\zeta}$$

$$+2B_2 e^{-r\zeta}[\sum_{k=1}^{\infty}\frac{(-r)^k}{k!(r+k)}e^{-k\zeta}\sum_{m=1}^{k}\frac{r(3r+3m-7)+(m-2)^2}{(r+m)(r+m-1)(r+m-2)} \quad (30)$$

$$+\zeta\sum_{k=0}^{\infty}\frac{(-r)^k}{k!(r+k)}e^{-k\zeta}]$$

Apart from removing the first summation in the last but-one-line of Appendix A, nothing else changes.

**Appendix D**

As velocity and temperature slip are inevitable in certain situations, and since fluid feed (suction or injection) through a porous surface is a used method of control of flow and heat transfer, we herein extend the exact solutions given above to include these effects. Following are the relevant changes. The subscript of the equation numbers refers to the corresponding equation of the main text.

For the velocity and magnetic field problem, the surface conditions in (7) are replaced by

$$f(0) = \psi, f'(0) = 1 + sf''(0) \quad (D1)_7$$

where $\psi$ represents the suction rate, and $s$ is the velocity slip coefficient.

The exact solutions have the forms

$$f = \psi + a(1-e^{-\gamma\eta}), \gamma > 0 \quad (D2)_{12}$$

$$g = be^{-\gamma\eta} \quad (D3)_{13}$$

where

$$a\gamma - a\psi - a^2 = \beta b \quad (D4)_{14}$$

$$b(\gamma - P_m\psi - P_m a) = a \tag{D5}_{15}$$

$$\gamma^2 as + \gamma a - 1 = 0 \tag{D6}$$

to satisfy Eqs. (2) and (3), and the boundary condition (D1b), respectively. The other boundary conditions are identically satisfied.

Equations (D4-6) manipulate to give the following polynomial equation for $\gamma$.

$$[s^2]\gamma^6 + [2s - (P_m + 1)\psi s^2]\gamma^5 + [1 - 2(P_m + 1)\psi s + (P_m\psi^2 - \beta)s^2]\gamma^4 \\ - [(P_m + 1)\psi + \{P_m(1 - 2\psi^2) + (1 + 2\beta)\}s]\gamma^3 \\ + [\{P_m(\psi^2 - 1) - (1 + \beta)\} + 2P_m\psi s]\gamma^2 + [2P_m\psi]\gamma + P_m = 0 \tag{D7a}_{16}$$

with

$$a = 1/(s\gamma^2 + \gamma) \tag{D7b}$$

and

$$b = a/(\gamma - P_m\psi - P_m a) \tag{D7c}$$

These solutions reduce to those obtained above when $\psi = s = 0$. Moreover, the solution for $f$ reduces to:

Gupta and Gupta's solution [5] when $s = 0$ so that $\gamma^2 - \psi\gamma - 1 = 0$, leading to $\gamma = \frac{1}{2}(\psi + \sqrt{\psi^2 + 4})$ and $a = \frac{1}{2}(-\psi + \sqrt{\psi^2 + 4})$,

Andersson's solution [6] when $\psi = 0$ so that $s\gamma^3 + \gamma^2 - 1 = 0$, leading to $a = \gamma$, and

Fang, Zhang and Yao's solution [8] when $P_m = 0$ so that $s\gamma^3 + (1 - \psi s)\gamma^2 - (s\beta + \psi)\gamma - (1 + \beta) = 0$, leading to $a = \gamma - \psi - \beta/\gamma$, $b = a/\gamma$

For the temperature constituents, the surface conditions (9a), (10a), and (11a) are replaced by

$$\theta_2(0) = \Theta_2 + \tilde{s}\theta_2'(0) \tag{D8}_{9a}$$

$$\theta_1(0) = \Theta_1 + \tilde{s}\theta_1'(0) \tag{D9}_{10a}$$

$$\theta_0(0) = \Theta_0 + \tilde{s}\theta_0'(0) \tag{D10}_{11a}$$

where $\tilde{s}$ is the thermal slip coefficient.

Upon substitution for $f$ and $g$, Eqs. (4), (5) and (6) become

$$\phi_2'' + r(\delta - e^{-\zeta})\phi_2' - 2re^{-\zeta}\phi_2 = -\tilde{\beta}a\gamma re^{-2\zeta} \tag{D11}_{20}$$

$$\phi_1'' + r(\delta - e^{-\zeta})\phi_1' - re^{-\zeta}\phi_1 = 0 \qquad (D12)_{20}$$

$$\phi_0'' + r(\delta - e^{-\zeta})\phi_0' = -2\phi_2 - 4a\gamma r e^{-2\zeta} \qquad (D13)_{20}$$

where the new parameters are $r = a\mathrm{Pr}/\gamma$, $\delta = \psi/a + 1$ and $\tilde{\beta} = \beta b^2/a^2 + 1$. Henceforth, it is assumed that $r\delta$ is not an integer.

The solutions are as follows.

$$\phi_2(\zeta) = -\tilde{\beta}a\gamma \frac{r}{2(2-r\delta)} e^{-2\zeta} + C_2 e^{-r\delta\zeta} M[-2 + r\delta, 1 + r\delta, -re^{-\zeta}] \qquad (D14)_{24}$$

$$\phi_1(\zeta) = C_1 e^{-r\delta\zeta} M[-1 + r\delta, 1 + r\delta, -re^{-\zeta}] \qquad (D15)_{27}$$

$$\begin{aligned}
\phi_0(\zeta) &= C_0 e^{-r\delta\zeta} M[r\delta, 1 + r\delta, -re^{-\zeta}] + a\gamma r \left(\tilde{\beta}\frac{1}{2-r\delta} - 4\right) \sum_{k=2}^{\infty} e^{-k\zeta} \frac{(-r)^{k-2}}{k(2-r\delta)^{(k-1)}} \\
&+ 2C_2 e^{-r\delta\zeta} \left[\sum_{k=1}^{\infty} e^{-k\zeta} \frac{(-r)^k}{k!(r\delta + k)} \sum_{m=1}^{k} \frac{r\delta(3r\delta + 3m - 7) + (m-2)^2}{(r\delta + m - 2)(r\delta + m - 1)(r\delta + m)} \right. \\
&\left. + \zeta \sum_{k=0}^{\infty} e^{-k\zeta} \frac{(-r)^k}{k!(r\delta + k)}\right]
\end{aligned} \qquad (D16)_{29}$$

The arbitrary constants of integration $C_n, n = 2, 1, 0$ are determined by the satisfaction of the surface conditions. For example, in the case of $n = 2$, condition (D8) gives the following equation for $C_2$

$$\begin{aligned}
-\tilde{\beta}a\gamma &\frac{r}{2(2-r\delta)} + C_2 M[-2 + r\delta, 1 + r\delta, -r] \\
&= \frac{\Theta_2}{\gamma^2} \\
&+ \tilde{\lambda}\gamma \left[\tilde{\beta}a\gamma \frac{r}{(2-r\delta)} + C_2\{-r\delta M[-2 + r\delta, 1 + r\delta, -r] \right. \\
&\left. + \frac{-2 + r\delta}{1 + r\delta} rM[-1 + r\delta, 2 + r\delta, -r]\}\right]
\end{aligned} \qquad (D17)_{25a}$$

while condition (9b) gives

$$\tilde{\beta}a\gamma\frac{r}{(2-r\delta)} + C_2\Big\{-r\delta M[-2+r\delta, 1+r\delta, -r]$$
$$+\frac{-2+r\delta}{1+r\delta}rM[-1+r\delta, 2+r\delta, -r]\Big\} = -\frac{Q_2}{\gamma^3} \qquad (D18)_{25b}$$

The convergence of the summations in Eq. (D16) can be established as before